\documentclass[12pt]{article}
\usepackage{texdraw}          
\begin{document}

\def\j{{\bf j}}
\def\tg{{ \rm tg}}

\begin{flushright} 
{hep-th/0410086}
\end{flushright}

\begin{center}

{\Large \bf Quantum  kinematics}
\end{center}

\begin{center}
{\Large N.A. Gromov, V.V. Kuratov} \\
 Department of Mathematics, \\
Syktyvkar Branch of IMM UrD RAS,  \\
Chernova st., 3a, Syktyvkar, 167982, Russia \\
E-mail: gromov@dm.komisc.ru
\end{center}

\begin{center}
{ \bf Abstract}
\end{center}

  The FRT quantum group and space theory is reformulated from the standard
mathematical basis to an arbitrary one. The $N$-dimensional quantum vector
Cayley-Klein spaces  are described in Cartesian basis and the quantum analogs 
of $(N-1)$-dimensional constant curvature spaces are introduced. Part of the 
$4$-dimensional constant curvature spaces are interpreted as the 
non-commutative analogs of $(1+3)$ kinematics. A different unifications of 
Cayley-Klein and Hopf structures in a kinematics are described with the help 
of permutations. All permutations which lead to the physically 
nonequivalent  kinematics are found and the corresponding non-commutative  
$(1+3)$ kinematics are investigated. As a result the quantum (anti) de Sitter, 
Minkowski, Newton, Galilei kinematics with the fundamental length,
the fundamental mass and the fundamental velocity are obtained.

\vspace{8mm}

\section{Introduction}

Space-time is a fundamental conception which underline the most significant
physical theories. Therefore the analysis of a possible space-time models
(or kinematics) has the fundamental  meaning  for physics. Space and time
in non-relativistic physics were regarded as independent what 
mathematically is connected with fiber property of Galilei kinematics.
In special relativity was determined that space and time are depend on 
each other and must be regarded as integrated object, namely flat Minkowski
space-time with pseudo-Euclidean metric.  The notion of curvature was
introduced in physics by general relativity. Anti de Sitter and de Sitter
kinematics with constant positive respectively negative curvature are
the simplest relativistic space-time models with curvature. Possible
kinematics, which satisfy the natural physical postulates:  space is
isotropic and rotations in space-time planes form non-compact subgroup,
were described in \cite{B-68} on the level of Lie algebras.

The Snyder quantized space-time coordinates \cite{S-47} or, respectively,
the curved momentum space is the oldest example of using the non-commutative
geometry in physics. The simplest curved de Sitter geometry with constant 
curvature were used instead of flat Minkowski space in different 
generalizations of quantum field theory \cite{K-61}--\cite{MK-96} as a
momentum space model. The universal constant, the fundamental length $l,$
or  fundamental mass $M,$ related to $l$ by $l={\frac {\hbar}{Mc}},$
where $\hbar$ is  Plank constant and $c$ velocity of light enters
necessarily into the theory \cite{K-61},\cite{T-65},\cite{MK-96}.

New possibility for construction of the non-commutative space-time models
is provided by quantum groups and quantum vector spaces \cite{FRT}.
Space-time coordinates commutation relations  of the  
$\kappa$-Minkowski kinematics \cite{Z-94}, \cite{KM-95}, \cite{LLM}
was obtained from Lie algebra quantum deformation of Poincar\'e group 
\begin{equation}
[x_{\mu},x_{\nu}]={\frac {i}{\kappa}}(a_{\mu}x_{\nu}-x_{\mu}a_{\nu}),
\label{i1}
\end{equation}
where $a_{\mu} $ is  four-vector in Minkowski space, determining  
the direction of quantum deformation $y=a^{\mu}x_{\mu}. $ 
The arbitrary choice of  $a_{\mu} $ is equivalent to the description of
standard $\kappa$-deformation in space-time with an arbitrary basis 
\cite{KM-95}.
If we put $\hbar=c=1 $ the deformation parameter 
$\Lambda=\kappa^{-1}=[\mbox{length}]$ may be treated
as the fundamental length parameter and $\kappa$ may be regarded as
the fundamental mass $[\kappa]=[\mbox{mass}].$

The standard  $\kappa$-deformation obtained for
$a_{\mu}=(1,0,0,0),\;a_{\mu}a^{\mu}=1 $  leads to relations $(i,k=1,2,3)$
\begin{equation}
[x_0,x_i]={\frac {i}{\kappa}}x_i,\quad [x_i,x_k]=0.
\label{i2}
\end{equation}
Space-like vector
$a_{\mu}=(0,1,0,0),\;a_{\mu}a^{\mu}=-1 $ 
define tachyonic $\kappa$-deformation $(p=2,3)$
\begin{equation}
[x_1,x_0]={\frac {i}{\kappa}}x_0,\; [x_p,x_0]=0,\;
[x_1,x_p]={\frac {i}{\kappa}}x_p,\; [x_2,x_3]=0.
\label{i3}
\end{equation}
Finally light-like vector
$a_{\mu}=(1,1,0,0),\;a_{\mu}a^{\mu}=0 $ 
provided light-cone $\kappa$-deformation
\begin{equation}
[x_0,x_1]={\frac {i}{\kappa}}(x_1-x_0),\; 
[x_0,x_p]={\frac {i}{\kappa}}x_p,\;
[x_1,x_p]={\frac {i}{\kappa}}x_p,\; 
[x_2,x_3]=0,
\label{i4}
\end{equation}
which was suggested in  \cite{BHOS-96} under the name of the null-plane
Poincar\'e algebra.

Possible commutative kinematics \cite{B-68} are realized \cite{GYa-86}
as a constant curvature spaces, which may be obtained from the spherical space
by contractions and analytical continuations  known as Cayley-Klein scheme
or Cayley-Klein structure \cite{G-90}. The standard quantum group theory
\cite{FRT} was reformulated to the Cartesian basis and the non-commutative
analogs of constant curvature spaces (CCS) including fiber (or flag) spaces
and their motion groups were investigated in \cite{GKK-97}--\cite{GK-97}.
Quantum algebras corresponding to the flat Cayley-Klein spaces were described
in \cite{BHOS-93}, \cite{BHOS-95}.

It was shown \cite{G-95} that Hopf algebra  and Cayley-Klein structures
for the quantum orthogonal algebra may be combined in a different way.
An arbitrary permutations of indices of its generators in Cartesian basis,
which transform a set of primitive generators to a new set of primitive 
generators, generally give in result the isomorphic quantum algebra.
But such isomorphism may be destroyed if a physical interpretation of
the generators is introduced or a contraction of quantum algebra is realized.
Contractions of the quantum orthogonal groups for arbitrary permutations
were investigated in  \cite{Sb-2000}--\cite{Sb-2003}.

In this paper $N$-dimensional quantum vector Cayley-Klein spaces are
regarded and quantum analogs of $(N-1)$-dimensional constant curvature
spaces are obtained. For $N=5$ some of the quantum CCS are interpreted
as non-commutative $(1+3)$ kinematics: (anti) de Sitter, Minkowski, Newton,
Galilei, as well as exotic Carroll ones. All permutations  
which corresponding to  physically different kinematics are found.

The paper is organized as follows. In section 2, the unified description
of the commutative CCS and their interpretation as kinematics are briefly
recalled. Section 3 is devoted to the reformulation of the standard quantum
group and space theory for an arbitrary basis and the description of the
$N$-dimensional quantum vector Cayley-Klein spaces in Cartesian basis. Non-commutative quantum
$(1+3)$ kinematics for different permutations are discussed in section 4.
The obtained results are summarized in Conclusion.

\section{Commutative kinematics}

Classical four-dimensional space-time models may be obtained
\cite{GYa-86}, \cite{G-90} by the physical interpretation
of the orthogonal  coordinates of the most symmetric spaces,
namely constant curvature spaces. All $3^N$ $N$-dimensional CCS
are realized on the spheres
\begin{equation}
S_N(j)=\{ \xi_1^2+j_1^2\xi_2^2+ \ldots +(1,N+1)^2\xi_{N+1}^2=1 \},
\label{c1}
\end{equation}
where 
\begin{equation}
(i,k)=\prod^{\max(i,k)-1}_{l=\min(i,k)}j_l, \quad
(k,k)\equiv 1,
\label{16}
\end{equation}
each of parameters  $j_k=1,\iota_k,i, \  k=1, \ldots, N.$  
Here $ {\iota}_k $ are nilpotent generators
 $  {\iota}_k^2=0,  $ with commutative law of multiplication
$ {\iota}_k{\iota}_m={\iota}_m{\iota}_k \not =0, \  k \neq m. $
Division of complex numbers on the nilpotent generators
$ a/{\iota}_k, \ a \in {\bf C}, $  as well as division of
nilpotent generators with different indices
$ {\iota}_k/{\iota}_p, \ k \not = p $  are not defined.
But it is possible consistently  define division of
nilpotent generators on itself, namely
$ {\iota}_k/{\iota}_k=1 $ means that an equation 
$ a{\iota}_k=b{\iota}_k $ has the single solution $ a=b$ for
$a,b \in {\bf R}$ or ${\bf C}.$

\begin{figure}[h]
\begin{texdraw}
\drawdim{mm}
\move(0 0)
\lvec(130 0) \lvec(130 90) \lvec(0 90)
\lvec(0 0) \move (10 0) \lvec(10 90)
\move(50 0) \lvec(50 90) \move(90 0) \lvec(90 90)
\move(0 40) \lvec(130 40)
\move(0 80) \lvec(130 80)
\textref h:C v:C \htext(5 20) {\footnotesize $i$}
\textref h:C v:C \htext(5 60) {\footnotesize $\iota_2$}
\textref h:C v:C \htext(30 85) {\footnotesize $1$}
\textref h:C v:C \htext(70 85) {\footnotesize $\iota_1$}
\textref h:C v:C \htext(110 85) {\footnotesize $i$}
\textref h:C v:C \htext(2.3 82.5) {\footnotesize $j_2$}
\textref h:C v:C \htext(8 87.5) {\footnotesize $j_1$}
\move(10 80) \linewd 0,5 \rlvec(-10 10)
%%%%%%%%%%%%%%%%%%%%%%%%%%%%%%
%%%%%%%%%%%%    ADS  %%%%%%%%%
%%%%%%%%%%%%%%%%%%%%%%%%%%%%%%
\move(30 5)\arrowheadtype t:V \arrowheadsize l:2 w:1 \linewd 0.2 \ravec(0 30)
\move(15 20)\ravec(30 0)
\move(46 20) \larc r:25 sd:155 ed:205
\move(14 20) \larc r:25 sd:-25 ed:25
\move(30 -12) \larc r:25 sd:63 ed:117
\move(30 52) \larc r:25 sd:243 ed:297
\move(30 20) \linewd 0.2 \lpatt(0.5 1) \rlvec(6 6)
\move(30 20) \rlvec(-6 -6) \move(30 20) \rlvec(6 -6) \move(30 20) \rlvec(-6 6) 
\lpatt()
\textref h:C v:C \htext(30 3) {\footnotesize anti de Sitter}
\textref h:C v:C \htext(33 35.5) {\footnotesize t}
\textref h:C v:C \htext(45 23) {\footnotesize r}
%%%%%%%%%%%%%%%%%%%%%%%%%%%%%%
%%%%%%%%%%%%%%  N(+) %%%%%%%%%
%%%%%%%%%%%%%%%%%%%%%%%%%%%%%%
\textref h:C v:C \htext(30 43) {\footnotesize Newton (+)}
\textref h:C v:C \htext(33 75.5) {\footnotesize t}
\textref h:C v:C \htext(45 63) {\footnotesize r}
\move(30 45)\linewd 0.2 \ravec(0 30)
\move(15 60)\linewd 0.5 \ravec(30 0)
\move(46 60) \linewd 0.2 \larc r:25 sd:155 ed:205
\move(14 60) \larc r:25 sd:-25 ed:25
\move(18 52)\linewd 0.5 \rlvec(24 0)
\move(18 68) \rlvec(24 0)
%%%%%%%%%%%%%%%%%%%%%%%%%%%%%%
%%%%%%%%%%%%   Min   %%%%%%%%%
%%%%%%%%%%%%%%%%%%%%%%%%%%%%%%
\move(70 5) \linewd 0.2 \ravec(0 30)
\move(55 20)\ravec(30 0)
\move(62 8) \rlvec(0 24)
\move(78 8) \rlvec(0 24)
\move(58 12) \rlvec(24 0)
\move(58 28) \rlvec(24 0)
\move(70 20) \linewd 0.2 \lpatt(0.5 1) \rlvec(6 6)
\move(70 20) \rlvec(-6 -6) \move(70 20) \rlvec(6 -6) \move(70 20) \rlvec(-6 6) 
\lpatt()
\textref h:C v:C \htext(70 3) {\footnotesize Minkowski}
\textref h:C v:C \htext(73 35.5) {\footnotesize t}
\textref h:C v:C \htext(85 23) {\footnotesize r}
%%%%%%%%%%%%%%%%%%%%%%%%%%%%%%
%%%%%%%%%%%%%   DS   %%%%%%%%%
%%%%%%%%%%%%%%%%%%%%%%%%%%%%%%
\move(110 5)\linewd 0.2 \ravec(0 30)
\move(95 20)\ravec(30 0)
\move(144 20) \larc r:25 sd:155 ed:205
\move(76 20) \larc r:25 sd:-25 ed:25
\move(110 4) \larc r:25 sd:60 ed:120
\move(110 36) \larc r:25 sd:240 ed:300
\move(110 20) \linewd 0.2 \lpatt(0.5 1) \rlvec(6 6)
\move(110 20) \rlvec(-6 -6) \move(110 20) \rlvec(6 -6) \move(110 20) \rlvec(-6 6) 
\lpatt()
\textref h:C v:C \htext(110 3) {\footnotesize de Sitter}
\textref h:C v:C \htext(113 35.5) {\footnotesize t}
\textref h:C v:C \htext(125 23) {\footnotesize r}
%%%%%%%%%%%%%%%%%%%%%%%%%%%%%%%
%%%%%%%%%%   Galilei %%%%%%%%%%
%%%%%%%%%%%%%%%%%%%%%%%%%%%%%%%
\textref h:C v:C \htext(70 43) {\small Galilei}
\textref h:C v:C \htext(73 75.5) {\footnotesize t}
\textref h:C v:C \htext(85 63) {\footnotesize r}
\move(70 45) \linewd 0.2 \ravec(0 30)
\move(55 60)\linewd 0.5 \ravec(30 0)
\move(58 52) \rlvec(24 0)
\move(58 68) \rlvec(24 0)
\move(62 48) \linewd 0.2 \rlvec(0 24)
\move(78 48) \rlvec(0 24)
%%%%%%%%%%%%%%%%%%%%%%%%%%%%%%
%%%%%%%%%%%%%  N(-)  %%%%%%%%%
%%%%%%%%%%%%%%%%%%%%%%%%%%%%%%
\textref h:C v:C \htext(110 43) {\footnotesize Newton (-)}
\textref h:C v:C \htext(113 75.5) {\footnotesize t}
\textref h:C v:C \htext(125 63) {\footnotesize r}
\move(110 45) \linewd 0.2 \ravec(0 30)
\move(95 60)\linewd 0.5 \ravec(30 0)
\move(144 60) \linewd 0.2 \larc r:25 sd:155 ed:205
\move(76 60) \larc r:25 sd:-25 ed:25
\move(98 52)\linewd 0.5 \rlvec(24 0)
\move(98 68) \rlvec(24 0)
\end{texdraw}
\caption{Classical (1+3) kinematics. Light cone is drawn by dotted lines.
 Proper space of the non-relativistic kinematics is represented by thick lines.}
\label{fig:Kinematics}
\end{figure}

The intrinsic Beltrami coordinates
$x_k=\xi_{k+1}\xi_1^{-1},\; k=1,2,\ldots,N $
present the  coordinate system on CCS, which coordinate lines
$x_k=const$ are geodesic. CCS has positive curvature for $j_1=1,$ 
negative for $j_1=i$ and are flat for  $j_1=\iota_1.$
For flat space the Beltrami coordinates  coincide with the Cartesian ones.
Nilpotent values $j_k=\iota_k, k>1$ correspond to a fiber (flag) spaces 
and imaginary values $j_k=i$ correspond to pseudo-Riemannian spaces.

Classical $(1+3)$ kinematics are obtained from CCS for
$N=4, \; j_1=1,\iota_1,i, \; j_2=\iota_2,i,\; j_3=j_4=1 $
if one interpret $x_1$ as the time axis $t=\xi_2\xi_1^{-1}$ and the rest
as the space axes $r_k=\xi_{k+2}\xi_1^{-1}, k=1,2,3.$
These kinematics are represented on Fig.1, where coordinate lines are drawn.

Except of standard  (anti) de Sitter, Minkowski, Newton, Galilei kinematics
theoretically are possible non-standard exotic Carroll kinematics
\cite{B-68},\cite{LL-65}, where space and time properties are changed
as compared with those of Galilei kinematics.
Time is absolute in Galilei kinematics: two simultaneous events in some
inertial reference frame remain simultaneous in any other one.
But in Carroll kinematics space is absolute: two events which take place
at  some space point of  inertial reference frame will take place
at the same space point in any other one. 
 
Carroll kinematics as well are realized  \cite{GYa-86}, \cite{G-90}
as CCS for $N=4, j_4=\iota_4, j_1=1,\iota_1,i, j_2=j_3=1,$
but with different physical interpretation of Beltrami coordinates, namely
$t=\xi_5 \xi_1^{-1}$ is time axis and 
$r_k=\xi_{k+1} \xi_1^{-1},\;k=1,2,3$ are space axes.
Carroll kinematics with flat proper space firstly described in  \cite{LL-65}
correspond to $j_1=\iota_1.$

\section{Quantum orthogonal groups and quantum  Cayley-Klein vector spaces}

According with FRT theory \cite{FRT} the algebra function on
quantum orthogonal group $ Fun(SO_q(N)) $ (or simply quantum orthogonal group $ SO_q(N) $) is the algebra of non-commutative polynomials of $ n^2 $
variables $ t_{ij}, i,j=1, \ldots, n, $ which are subject of commutation
relations
\begin{equation}
R_qT_1T_2=T_2T_1R_q
\label{1}
\end{equation}
and additional relations of $q$-orthogonality
\begin{equation}
TCT^t=C,\quad T^tC^{-1}T=C^{-1}.
\label{2}
\end{equation}
Here
 $ T_1=T \otimes I, \ 
 T_2=I \otimes T \in M_{n^2}({\bf C} \langle t_{ij} \rangle), $
$ T=(t_{ij})^n_{i,j=1} \in M_n({\bf C} \langle t_{ij} \rangle ), $ 
 $ I $ is unit matrix in  $ M_n({\bf C}), $ 
 $ C=C_0q^{\rho}, $  $ \rho=diag(\rho_1, \ldots, \rho_N), \
(C_0)_{ij}=\delta_{i'j}, \ i'=N+1-i, \  i,j=1, \ldots, N, $ that is
 $ (C)_{ij}=q^{\rho_{i'}}\delta_{i'j} $  and
$ C^{-1}=C, $
\begin{equation}
(\rho_1, \ldots, \rho_N)=
\left \{ \begin{array}{ccc}
     (n-\frac{1}{2}, n-\frac{3}{2}, \ldots , \frac{1}{2},0,-\frac{1}{2},
     \ldots , -n+\frac{1}{2}), \; N=2n+1 \\
     (n-1, n-2, \ldots, 1,0,0,-1, \ldots, -n+1), \; N=2n.
     \end{array} \right.
\label{q8q}
\end{equation}
The numerical matrix $ R_q $ is the well-known solution \cite{FRT}
of Yang-Baxter equation and its elements serve as the structure constant
of quantum group generators.

Quantum orthogonal group $ SO_{q}(N) $ is a Hopf algebra with the
following coproduct   $\Delta,$  counit $ \epsilon $ and antipode $S$
$$
\Delta T=T \dot {\otimes} T, \quad
\Delta t_{ij}=\sum^n_{k=1} t_{ik} \otimes t_{kj},\quad
\epsilon(T)=I, \ \epsilon(t_{ij})=\delta_{ij},
$$
\begin{equation}
S(T)=CT^tC^{-1}, \quad  S(t_{ij})=q^{\rho_{i'}-\rho_{j'}}t_{j'i'}, \;
i,j=1, \ldots, N.
\label{6}
\end{equation}
  
Let us remind the definition of the quantum vector space \cite{FRT}.
An algebra  $ O_q^N({\bf C}) $ with generators $ x_1,\ldots,x_N $ and
commutation relations
\begin{equation}
\hat R_q(x \otimes x)=qx \otimes x-{{q-q^{-1}} \over {1+q^{N-2}}}
x^tCx  W_q,
\label{7}
\end{equation}
where
$ \hat R_q=PR_q, \; Pu \otimes v=v \otimes u,\;\forall u,v \in {\bf C}^n,\;
W_q=\sum^N_{i=1}q^{\rho_{i'}}e_i \otimes e_{i'},$ 
\begin{equation}  
x^tCx=\sum^N_{i,j=1}x_iC_{ij}x_j=
\epsilon x_{n+1}^2+
\sum_{k=1}^n\left(
q^{-\rho_k}x_kx_{k'}+q^{\rho_k}x_{k'}x_k\right), 
\label{8}
\end{equation}
$\epsilon =1 $ for $N=2n+1,$ $\epsilon =0 $ for $N=2n$
and vector $ (e_i)_k = \delta_{ik}, \enskip i,k=1,\ldots,N $
is called the algebra of functions on $N$-dimensional quantum
Euclidean space (or simply the quantum vector space) $ O_q^N({\bf C}). $

Co-action of the quantum group $ SO_{q}(N) $ on the non-commutative
vector space $ O_q^N({\bf C}) $ is given by
\begin{equation}
\delta(x)=T \dot \otimes x, \quad
\delta(x_i)=\sum^n_{k=1}t_{ik} \otimes x_k, \;  i=1, \ldots,n
\label{9}
\end{equation}
and quadratic form  (\ref{8}) is invariant $inv=x^tCx$  with respect to
this co-action:
\begin{equation}
\delta(x^tCx)=I  \otimes x^tCx. 
\label{10}
\end{equation}

The matrix $C$ has non-zero elements only on the secondary diagonal which are equal to unit in commutative limit $q=1$. Therefore the quantum group
$ SO_q(N) $ and the quantum vector space $ O_q^N({\bf C}) $ are described
by equations (\ref{1}),(\ref{2}),(\ref{7}),(\ref{8}) in mathematical 
(or ``symplectic'') basis, where for $q=1$ the invariant form $inv=x^tC_0x$  is given by the matrix $C_0$ with the only non-zero elements  on the secondary diagonal which are equal to real units.

New generators $y=D^{-1}x$ of the vector space $ O_q^N({\bf C}) $
in {\it a arbitrary} basis are obtained \cite{GKK-97}--\cite{GK-97}
with the help of non-degenerate matrix $D \in M_N$ and are subject of
the commutation relations
\begin{equation}
\hat {R}(y \otimes y)=
qy \otimes y - {\lambda \over 1+q^{N-2}} y^tC'y W,
\label{11}
\end{equation}
where $ {\hat {R}}=(D \otimes D)^{-1}{\hat R_q}(D \otimes D),\;
W=(D \otimes D)^{-1}W_q,\; C'=D^tCD.$
The corresponding quantum group $ SO_q(N) $ is generated in arbitrary
basis by $ U=(u_{ij})^N_{i,j=1}, $ where  $ U=D^{-1}TD. $
The commutation relations of the new generators are
\begin{equation}
 \tilde{R}U_{1}U_{2} = U_{2}U_{1} \tilde{R}
\label{12}
\end{equation}
and $q$-orthogonality relations looks as follows
\begin{equation}
U\tilde {C}U^t=\tilde {C}, \quad U^t(\tilde {C})^{-1}U=(\tilde {C})^{-1},
\label{13}
\end{equation}
where $ {\tilde {R}}=(D \otimes D)^{-1}{ R_q}(D \otimes D),\;
\tilde {C}=D^{-1}C(D^{-1})^t. $

In the case of kinematics most natural is the Cartesian basis where
the invariant form $inv=y^ty$ is given by the unit matrix $I.$
The transformation from the mathematical (or ``symplectic'')  basis $x$
to the Cartesian basis $y$ is described by the matrix $D$ which is a 
solution of the following equation
\begin{equation}
D^tC_0D=I.
\label{14}
\end{equation}
This equation has many solutions. Take one of these, namely
\begin{equation}
D=\frac{1}{\sqrt{2}}
\left ( \begin{array}{ccc}
      I & 0 &  -i{\tilde C_0} \\
      0 & \sqrt{2} &  0 \\
      {\tilde C_0} & 0  &  iI
      \end{array} \right ),    \       N=2n+1,
\label{15}
\end{equation}
where $ {\tilde C_0} $ is the  $ n \times n $ matrix  with real units
on the secondary diagonal. For $N=2n$ the matrix $D$ is given by (\ref{15})
without the middle column and row. All solutions of (\ref{14}) are given
by  the matrices $ D_{\sigma}=DV_{\sigma}, $ which are obtained from
(\ref{15}) by the right multiplication on the matrix $ V_{\sigma} \in M_{N} $ with elements
$ (V_{\sigma})_{ik}= \delta_{\sigma_{i},k}, $
where $ \sigma \in S(N) $ is a permutation of $ N$-th order
\cite{Sb-2000}--\cite{Sb-2003}.

We derive the quantum Cayley-Klein spaces with the same  transformation
of Cartesian generators
$ y =\psi \xi, \; \psi ={\rm diag}(1,(1,2),\ldots,(1,N)) \in M_N, $
as in commutative case \cite{G-90}, \cite{G-97}.
The transformation  $z=Jv$ of the deformation parameter $q=e^z$ need be added in quantum case. The commutation relations of the Cartesian generators of the quantum  $N$-dimensional Cayley-Klein space are given by equations  
$$
{{\hat R}}_{\sigma}(j) \xi \otimes \xi = e^{Jv} \xi \otimes \xi -
{{2sh Jv} \over {1+e^{Jv(N-2)}}}\xi^tC_{\sigma}(j) \xi  {W}_{\sigma}(j),
$$
$$
{\hat R}_{\sigma}(j)={\Psi}^{-1}{\hat R}_{\sigma}\Psi, \quad
W_{\sigma}(j)={\Psi}^{-1}{W}_{\sigma},
$$
\begin{equation}
C_{\sigma}(j)=
{\psi} D^t_{\sigma} C D_{\sigma} \psi =
{\psi} V^{t}_{\sigma} D^t C D V_{\sigma} \psi,\;\;
\Psi = \psi \otimes \psi
\label{17}
\end{equation}
and in explicit form are
      $$
\xi_{\sigma_{k}}\xi_{\sigma_{m}}  =  
\xi_{\sigma_{m}} \xi_{\sigma_{k}} \cosh Jv - 
i\xi_{\sigma_{m}} \xi_{\sigma_{k'}}
(1,{\sigma_{k'}})(1,{\sigma_{k}})^{-1}\sinh Jv, \;
k<m<{k'}, \;  k \not = m',
      $$
      $$
\xi_{\sigma_{k}}\xi_{\sigma_{m}}  =  
\xi_{\sigma_{m}} \xi_{\sigma_{k}} \cosh Jv - 
i\xi_{\sigma_{m'}} \xi_{\sigma_{k}} 
(1,{\sigma_{m'}})(1,{\sigma_{m}})^{-1} \sinh Jv, \;
m'<k<m, \;  k \not = m',
      $$
      $$ 
 [ \xi_{\sigma_{k}},\xi_{\sigma_{k'}}  ]  
=  2i\epsilon
\sinh({{Jv} \over {2}}){(\cosh Jv)}^{n-k} \xi^2_{\sigma_{n+1}}
 \displaystyle{ {(1,{\sigma_{n+1}})^{2}}\over
{(1,{\sigma_{k}})(1,{\sigma_{k'}}) } } + 
      $$
\begin{equation}       
+ i\displaystyle{ {\sinh(Jv)} \over{(\cosh Jv)}^{k+1}
(1,{\sigma_{k}})(1,{\sigma_{k'}}) }
    \sum^n_{m=k+1}
   {(\cosh   Jv)}^m  ((1,{\sigma_{m}})^{2}\xi^2_{\sigma_{m}} + 
(1,{\sigma_{m'}})^{2} \xi^2_{\sigma_{m'}}), 
\label{18}
 \end{equation}
where $k,m=1,2,\ldots,n. $ 
The invariant form of the Cayley-Klein space 
$ O^N_v(j;\sigma;{\bf C}) $ is written as
$$
inv(j) =
\cosh (Jv{ \rho_1})\Biggl(  \epsilon (1,\sigma_{n+1})^2\xi^2_{\sigma_{n+1}}
{{(\cosh Jv)^{n}} \over {\cosh(Jv/2)}} +
$$
\begin{equation}
+ \sum^n_{k=1} ( (1,\sigma_{k})^{2}\xi^2_{\sigma_{k}} +
(1,\sigma_{k'})^{2}\xi^2_{\sigma_{k'}}) {(\cosh Jv)}^{k-1}\Biggr).
\label{19}
\end{equation}
The multiplier $J$ in the transformation  $z=Jv$ of the deformation parameter is chosen as
$ J=\displaystyle{\bigcup^n_{k=1}(\sigma_k,\sigma_{k'})}.$
This is the minimal multiplier, which guarantees \cite{Sb-2003}
the existence of the Hopf algebra structure for the associated quantum group $SO_v(N;j;\sigma).$ 
The ``union''
$(\sigma_k,\sigma_{p}) \bigcup  (\sigma_m,\sigma_{r})$
is understood as the first power multiplication of all parameters
$j_k,$ which are appear at least in one multiplier
$(\sigma_k,\sigma_{p})  $ or $(\sigma_m,\sigma_{r}),  $
for example, $(j_1j_2) \bigcup (j_2j_3)=j_1j_2j_3.$

In the case of Euclidean vector spaces $ O_q^N({\bf C}) $ the use
of different $D_{\sigma}$ for $\sigma\in S(N)$ has no sense because
all quantum vector spaces are isomorphic. But the matter is radically different 
for the quantum Cayley-Klein spaces. In this case Cartesian generators 
are multiplied by $(1,k)$ and for nilpotent values of all or
some parameters $j_k$ such isomorphism of quantum vector spaces is  destroyed.
 The necessity of using different $ D_{\sigma} $ is arisen
as well if there is some physical interpretation of generators.
In this case a physically different generators may be confused by
permutations $ \sigma, $ for example, time and space generators of kinematics.
Mathematically isomorphic kinematics may be physically non-equivalent.

 \section{ Quantum vector spaces $O^5_v(\j;\sigma)$}

Quantum vector spaces $O^5_v(\j;\sigma)$ are generated by
$\xi_{\sigma_l},\; l=1,2,3,4,5,$ with commutation relations $(k=2,3,4)$
      $$
\xi_{\sigma_1}\xi_{\sigma_k} = \xi_{\sigma_k}\xi_{\sigma_1} \cosh(Jv) 
-i\xi_{\sigma_k}\xi_{\sigma_5}{{(1,\sigma_5)} \over {(1,\sigma_1)}} \sinh(Jv), \;
     $$
     $$
\xi_{\sigma_k}\xi_{\sigma_5} = \xi_{\sigma_5}\xi_{\sigma_k} \cosh(Jv) 
 - i\xi_{\sigma_1}\xi_{\sigma_k}{{(1,\sigma_1)} \over {(1,\sigma_5)}} \sinh(Jv),
\;\; 
      $$
      $$
\xi_{\sigma_2}\xi_{\sigma_3} = \xi_{\sigma_3}\xi_{\sigma_2} \cosh(Jv) 
-i\xi_{\sigma_3}\xi_{\sigma_4}{{(1,\sigma_4)} \over {(1,\sigma_2)}} \sinh(Jv), \;
      $$
      $$
\xi_{\sigma_3}\xi_{\sigma_4} = \xi_{\sigma_4}\xi_{\sigma_3} \cosh(Jv) 
 - i\xi_{\sigma_2}\xi_{\sigma_3}{{(1,\sigma_2)} \over {(1,\sigma_4)}} \sinh(Jv),
      $$
      $$
\left  [\xi_{\sigma_2},\xi_{\sigma_4}  \right  ]  =
2i\xi_{\sigma_3}^2{{(1,\sigma_3)^2} \over {(1,\sigma_2)(1,\sigma_4)}} \sinh(Jv/2),
      $$
$$
 [ \xi_{\sigma_1},\xi_{\sigma_5}  ] = 
2i\Biggl( \xi_{\sigma_3}^2 (1,\sigma_3)^2 \cosh(Jv) +
( \xi_{\sigma_2}^2 (1,\sigma_2)^2 + 
$$
\begin{equation}
+\xi_{\sigma_4}^2 (1,\sigma_4)^2 )\cosh(Jv/2)  \Biggr)
{{\sinh(Jv/2)}\over {(1,\sigma_1)(1,\sigma_5)}}.
\label{1q}
\end{equation}
Coaction of $SO_v(5;j;\sigma) $ on $O_v^5(j;\sigma)$ is given by
\begin{equation}
\delta(\xi(j;\sigma))=U(j;\sigma)\dot \otimes \xi(j;\sigma)
\label{qq2}
\end{equation}
and the following form
    $$
inv(j)  =  \Biggl( \xi_{\sigma_3}^2 (1,\sigma_3)^2 
{{(\cosh(Jv))^2}\over {\cosh(Jv/2)}} +  
\xi_{\sigma_1}^2 (1,\sigma_1)^2 +   
\xi_{\sigma_5}^2 (1,\sigma_5)^2 +   
    $$
\begin{equation}
+( \xi_{\sigma_2}^2(1,\sigma_2)^2  +
 \xi_{\sigma_4}^2(1,\sigma_4)^2  )\cosh(Jv)   
\Biggr) \cosh(3Jv/2)
\label{2q}
\end{equation}
is invariant under this coaction.

Quantum orthogonal Cayley-Klein sphere $S_v^4(j;\sigma) $ is obtained
as the quotient of $O_v^5(j;\sigma) $ by $ inv(j)=1. $ 
Generators 
$$
\zeta_{\sigma_1}=Au_{\sigma_1\sigma_k}, \;
\zeta_{\sigma_2}=Au_{\sigma_2\sigma_k}, \;
\zeta_{\sigma_3}=Au_{\sigma_3\sigma_k}, \;
$$
\begin{equation}
\zeta_{\sigma_4}=Au_{\sigma_4\sigma_k}, \;
\zeta_{\sigma_5}=Au_{\sigma_5\sigma_k}, \;
\sigma_k=1
\label{3q}
\end{equation}
of the quantum orthogonal sphere $S^4_v(j,\sigma),$  forming the  vector
$$
\zeta^t(j;\sigma)=
((1,\sigma_1)\zeta_{\sigma_1}, (1,\sigma_2)\zeta_{\sigma_2},
(1,\sigma_3)\zeta_{\sigma_3})^t, (1,\sigma_4)\zeta_{\sigma_4},
(1,\sigma_5)\zeta_{\sigma_5}),
$$
are proportional to the first column elements of the matrix $U(j;\sigma). $
It follows from the $(v,j)$-orthogonality of $U(j,\sigma),$ that 
the additional relation 
\begin{equation}
\zeta^t(j;\sigma)C(j)\zeta(j;\sigma)=1
\label{4q}
\end{equation}
is held for generators therefore they belong to $S^4_v(j;\sigma).$
The quantum analogs of the intrinsic Beltrami coordinates on this sphere
are given by  the set of independent generators
\begin{equation}
x_{\sigma_{i-1}}=\zeta_{\sigma_i}\cdot \zeta_1^{-1},
\quad i=1,2,3,4,5, \quad i\neq k.
\label{5q}
\end{equation}

The systematic investigation of the  quantum vector spaces
and the quantum orthogonal spheres for different $\sigma $ and different contractions give in result a quantum analogs of $(1+3)$ kinematics.
It is easily to find out that the commutation relations are invariant
relative to the replacement of $ \sigma_1$ on $ \sigma_5,$ $ \sigma_2$ on $ \sigma_4$ and vice-versa, therefore it is sufficient to consider only
30 permutations instead of 5!=120. Moreover only 3 permutations 
$ \sigma_0=(1,2,3,4,5),\;  \sigma'=(1,4,3,5,2),\; \tilde{\sigma}=(2,3,1,4,5) $ 
give in result non-isomorphic quantum kinematics.  

%%%%%%%%%%%%%%%%%%%%%%%%%%%%%%%%%%%%%%%%%%%%%%%%%%%%%%
\subsection{ Quantum vector space $O^5_v(\j;\sigma_0)$}
%%%%%%%%%%%%%%%%%%%%%%%%%%%%%%%%%%%%%%%%%%%%%%%%%%%%%
For the identical permutation  $ \sigma_0=(1,2,3,4,5)  $
deformation parameter is multiplied by $J=(1,5)=j_1j_2j_3j_4$
and commutation relations of the generators are  $(m=2,3,4) $
 $$
\xi_{1}\xi_{m}=\xi_{m}\xi_{1} \cosh Jv
-i\xi_{m}\xi_{5} J \sinh Jv , \quad
\xi_{m}\xi_{5}=\xi_{5}\xi_{m} \cosh Jv
-i\xi_{1}\xi_{m} {1 \over J }\sinh Jv , 
$$
$$
\xi_{2}\xi_{3}=\xi_{3}\xi_{2} \cosh Jv
-i\xi_{3}\xi_{4} j_2j_3 \sinh Jv , \quad
\xi_{3}\xi_{4}=\xi_{4}\xi_{3} \cosh Jv
-i\xi_{2}\xi_{3} {1 \over j_2j_3 }\sinh Jv ,
$$
$$
 [\xi_{1},\xi_{_5}] = 
2ij_1^2 \Biggl( j_2^2\xi_{3}^2  \cosh(Jv) +
( \xi_{2}^2  + 
j_2^2j_3^2\xi_{4}^2  )\cosh{\frac {Jv}{2}} \Biggr)
{ 1 \over J}\sinh{\frac {Jv}{2}}, 
$$
\begin{equation}
[\xi_{2},\xi_{4}]=2i\xi_3^2{j_2 \over j_3 }\sinh{\frac {Jv}{2}}.
\label{6q}
\end{equation}
The following quadratic form 
$$
inv(j)  =  \Biggl( 
\xi_{1}^2  + J^2 \xi_{5}^2  +   
+ j_1^2 ( \xi_{2}^2  + j_2^2j_3^2 \xi_{4}^2  )\cosh(Jv)+ 
$$
\begin{equation}
 + j_1^2j_2^2\xi_{3}^2 {{(\cosh(Jv))^2}\over {\cosh({Jv/2})}} 
\Biggr) \cosh{\frac {3Jv}{2}}
\label{7q}
\end{equation}
is invariant under the coaction of $SO_v(5;j;\sigma_0).$

For the standard interpretation of the independent generators
$t=\xi_2\xi_1^{-1},\;$ $ r_k=\xi_{k+2}\xi_1^{-1},\; k=1,2,3,$
which give rise to the (anti) de Sitter, Minkowski, Newton and Galilei
kinematics \cite{GYa-86}, \cite{G-90}, the mathematical parameter $j_1$
is replaced by  the physical one  $\tilde{j}_1T^{-1},$ 
and the  parameter  $j_2$ is replaced by  the $ic^{-1},$ where $\tilde{j}_1=1,i.$ 
The limit $T \rightarrow \infty $ correspond to the contraction  $j_1=\iota_1$ and the limit $c \rightarrow \infty $ correspond to
$j_2=\iota_2.$ In result the commutation relations (\ref{6q}) are
rewritten as follows $(m=2,3,4) $
$$
\xi_{1}\xi_{m}=\xi_{m}\xi_{1} \cos {\frac{\tilde{j}_1v}{cT}}
+i \xi_m\xi_5 {\frac{\tilde{j}_1}{cT}} \sin {\frac{\tilde{j}_1v}{cT}}, 
$$
$$
\xi_{m}\xi_{5}=\xi_{5}\xi_{m} \cos {\frac{\tilde{j}_1v}{cT}}
-i \xi_1\xi_m {\frac{cT}{\tilde{j}_1}} \sin {\frac{\tilde{j}_1v}{cT}}, 
$$
$$
\xi_{2}\xi_{3}=\xi_{3}\xi_{2} \cos {\frac{\tilde{j}_1v}{cT}}
+i \xi_3\xi_4 {\frac{1}{c}} \sin {\frac{\tilde{j}_1v}{cT}}, 
$$
$$
\xi_{3}\xi_{4}=\xi_{4}\xi_{3} \cos {\frac{\tilde{j}_1v}{cT}}
-i \xi_2\xi_3 c \sin {\frac{\tilde{j}_1v}{cT}}, 
$$
$$
[\xi_{1},\xi_{5}]= 2i {\frac{\tilde{j}_1c}{T}} \left (
(\xi_2^2 - {\frac{1}{c^2}} \xi_4^2)  \cos {\frac{\tilde{j}_1v}{2cT}}-
{\frac{1}{c^2}}\xi_3^2 \cos {\frac{\tilde{j}_1v}{cT}} \right )
\sin {\frac{\tilde{j}_1v}{2cT}},
$$
\begin{equation}
[\xi_{2},\xi_{4}]= -2i\xi_3^2
 {\frac{1}{c}} \sin {\frac{\tilde{j}_1v}{2cT}}.
\label{8q}
\end{equation}

As far as the generator $\xi_1$ do not commute with $\xi_s,\;s=2,3,4,5$
it is  convenient to introduce right and left time
$t=\xi_2\xi_1^{-1}, \hat{t}=\xi_1^{-1}\xi_2$ and space
$r_k=\xi_{k+2}\xi_1^{-1}, \hat{r}_k=\xi_1^{-1}\xi_{k+2}, k=1,2,3$
generators. The reason for such definition is the simplification of expressions for commutation relations of the (anti) de Sitter
quantum kinematics. It is possible to use only say rihgt generators,
but its commutators  are cumbersome.
The commutation relations of the independent generators are obtained
from (\ref{8q}) in the form
$$
S_v^{4(\pm)}(\sigma_0)=\{t,{\bf r}|\;\;
\hat{t}r_1=\hat{r}_1t \cos {\frac{\tilde{j}_1v}{cT}}
+i \hat{r}_1r_2 {\frac{1}{c}} \sin {\frac{\tilde{j}_1v}{cT}}, \;
$$
$$
\hat{t}r_2 - \hat{r}_2t= -2i\hat{r}_1r_1{\frac{1}{c}}  
\sin {\frac{\tilde{j}_1v}{2cT}},\;\;
\hat{t}r_3=\hat{r}_3t \cos {\frac{\tilde{j}_1v}{cT}}
-i t {\frac{cT}{\tilde{j}_1}} \sin {\frac{\tilde{j}_1v}{cT}}, \;
$$
$$
\hat{r}_1r_2=\hat{r}_2r_1 \cos {\frac{\tilde{j}_1v}{cT}}
-i \hat{t}r_1 c \sin {\frac{\tilde{j}_1v}{cT}}, \;\;
$$
\begin{equation}
\hat{r}_pr_3=\hat{r}_3r_p \cos {\frac{\tilde{j}_1v}{cT}}
-i r_p {\frac{cT}{\tilde{j}_1}} \sin {\frac{\tilde{j}_1v}{cT}} \}.
\label{9q}
\end{equation}

 The right and left generators are connected as follows 
$$
r_3-\hat{r}_3=2i{\frac{\tilde{j}_1}{cT}} \left (
(\hat{t}t-{\frac{1}{c^2}} \hat{r}_2r_2) \cos {\frac{\tilde{j}_1v}{2cT}}
-i{\frac{1}{c^2}}\hat{r}_1r_1 \cos {\frac{\tilde{j}_1v}{cT}} \right )
 \sin {\frac{\tilde{j}_1v}{2cT}}, 
$$
$$
\hat{r}_p=r_p \cos {\frac{\tilde{j}_1v}{cT}}
-i \hat{r}_3r_p {\frac{\tilde{j}_1}{cT}} \sin {\frac{\tilde{j}_1v}{cT}},\;p=1,2,
$$
\begin{equation}
\hat{t}=t \cos {\frac{\tilde{j}_1v}{cT}}
-i \hat{r}_2t {\frac{\tilde{j}_1}{cT}} \sin {\frac{\tilde{j}_1v}{cT}}.
\label{9q-1}
\end{equation}

The commutation relations of the time $t,\hat{t}$ and space $r_k,\hat{r}_k$
generators of the $(1+3)$ non-commutative quantum de Sitter $S_v^{4(-)}(\sigma_0)$ and anti de Sitter $S_v^{4(+)}(\sigma_0)$
kinematics are given by (\ref{9q}) for $\tilde{j}_1=i$ and
$\tilde{j}_1=1,$ respectively. The parameter $T$ is interpreted as
the curvature radius and has the time physical dimension
$[T]=[\mbox{time}],$ the parameter  $c$ is the light velocity  
$[c]=[\mbox{length}][\mbox{time}]^{-1}, $ deformation parameter
$v$ for the system units, where $\hbar=1,$ has the physical dimension
of length $[v]=[cT]=[\mbox{length}]=[\mbox{momentum}]^{-1} $ 
and may be interpreted as the fundamental length.

The quantum $(1+3)$ Minkowski kinematics $M_v^4(\sigma_0)$ is obtained
from  the quantum (anti) de Sitter kinematics $S_v^{4(\pm)}(\sigma_0)$
in the zero curvature limit  $T \rightarrow \infty. $ 
The left and right generators are the same $\hat{t}=t, \hat{r}_k=r_k$ 
in this limit and we have
$$
M_v^4(\sigma_0)=\{t,{\bf r}|\;\;  [t,r_p]=0, \;[r_3,t]=ivt, \; 
$$
\begin{equation}
[r_1,r_2]=0,\; [r_3,r_p]=ivr_p, \;p=1,2    \}.
\label{10q}
\end{equation}
This kinematics is isomorphic to the tachyonic $\kappa$-deformation of the
Minkowski kinematics (\ref{i3}), where $v=\Lambda=\kappa^{-1}$.

In the non-relativistic limit $ c \rightarrow \infty $ the quantum 
kinematics $S_v^{4(\pm)}(\sigma_0)$ are contracted to the non-commutative
analogs of $(1+3)$ Newton kinematics
 $N_v^{4(\pm)}(\sigma_0)$ with non-zero curvature.
In this limit $\hat{t}=t,\; \hat{r}_p=r_p,\; 
\hat{r}_3=r_3 -iv\tilde{j}_1^2t^2 / T^2  $ and the commutation relations
of the right space and time generators are as follows
$$
N_v^{4(\pm)}(\sigma_0)=\{t,{\bf r} | \; [t,r_p]=0, \;\;
[r_3,t]=ivt(1+\tilde{j}_1^2{\frac{t^2}{T^2}}), \;\;
$$        
\begin{equation}
[r_1,r_2]=0,\;\;
[r_3,r_p]=ivr_p(1+\tilde{j}_1^2{\frac{t^2}{T^2}}),\; p=1,2 \}. 
\label{11q}
\end{equation}
 
In the zero curvature limit $T \rightarrow \infty $
both quantum Newton kinematics are passed into the quantum Galilei
kinematics
$$
G_v^4(\sigma_0)=\{t,{\bf r}|\;\;  [t,r_p]=0, \;
[r_3,t]=ivt, 
$$
\begin{equation}
[r_1,r_2]=0,\;\; [r_3,r_p]=ivr_p,\; p=1,2 \},
\label{12q}
\end{equation}
which commutation relations are identical with those of the quantum
Minkowski kinematics (\ref{10q}).
 
Carroll kinematics  \cite{B-68}, \cite{LL-65} are also realized as constant curvature spaces, but with different interpretation of the
Beltrami coordinates, namely $r_k=\xi_{k+1} \xi_1^{-1},\;k=1,2,3$ 
are  the space generators and $t=\xi_5 \xi_1^{-1}$ is the time generator
\cite{GYa-86}, \cite{G-90}. 
Due to this  interpretation the new physical  dimensions of the contraction parameters are appeared: the parameter $j_1$ is replaced by
$\tilde{j}_1R^{-1},$ where $R \rightarrow \infty $ correspond to 
$j_1=\iota_1$ and $[R]=[\mbox{length}]; $ the parameter $j_4$ is replaced by $c,$ where $c \rightarrow 0 $ correspond to $j_4=\iota_4$ and
 $[c]=[\mbox{velocity}]. $
The deformation parameter
$[v]=[R][c]^{-1}=[\mbox{time}]=[\mbox{energy}]^{-1}$
is interpreted as the fundamental time.
The commutation relations (\ref{6q}) are rewritten in the form $(m=2,3,4)$
$$
\xi_{1}\xi_{m}=\xi_{m}\xi_{1} \cosh {\frac{\tilde{j}_1cv}{R}}
-i \xi_m\xi_5 {\frac{\tilde{j}_1c}{R}} \sinh {\frac{\tilde{j}_1cv}{R}}, \;\;
$$
$$
\xi_{m}\xi_{5}=\xi_{5}\xi_{m} \cosh {\frac{\tilde{j}_1cv}{R}}
-i \xi_1\xi_m {\frac{R}{\tilde{j}_1c}} \sinh {\frac{\tilde{j}_1cv}{R}},\;\;
$$
$$
\xi_{2}\xi_{3}=\xi_{3}\xi_{2} \cosh {\frac{\tilde{j}_1cv}{R}}
-i \xi_3\xi_4  \sinh {\frac{\tilde{j}_1cv}{R}},\;\;
$$
$$
\xi_{3}\xi_{4}=\xi_{4}\xi_{3} \cosh {\frac{\tilde{j}_1cv}{R}}
-i \xi_2\xi_3  \sinh {\frac{\tilde{j}_1cv}{R}},\;\;
$$
$$
[\xi_{1},\xi_{5}]= 2i {\frac{\tilde{j}_1}{cR}} \left (
(\xi_2^2+\xi_4^2)\cosh {\frac{\tilde{j}_1cv}{2R}}
+\xi_3^2  \cosh {\frac{\tilde{j}_1cv}{R}}   \right )
\sinh {\frac{\tilde{j}_1cv}{2R}},
$$
\begin{equation}
[\xi_{2},\xi_{4}]= 2i\xi_3^2 \sinh {\frac{\tilde{j}_1cv}{2R}}
\label{13q}
\end{equation}
and in the limit  $c \rightarrow 0 $ are
$$
[\xi_{1},\xi_{m}]=0,\;\; [\xi_{2},\xi_{3}]=0,\;\;  [\xi_{2},\xi_{4}]=0,\;\;
[\xi_{3},\xi_{4}]=0,\;\;
$$
\begin{equation}
\xi_{m}\xi_{5}=\xi_{5}\xi_{m}-iv\xi_{1}\xi_{m},\quad
[\xi_{1},\xi_{5}]= iv{\frac{\tilde{j}_1^2}{R^2}} (\xi_2^2+\xi_3^2+\xi_4^2).
\label{14q}
\end{equation}
Introducing  space $r_k=\xi_{k+1} \xi_1^{-1}, \; k=1,2,3$ and time
$t=\xi_{5} \xi_1^{-1},\;\hat{t}=\xi_1^{-1} \xi_5$ generators and 
taking into account that 
$\hat{t}=t- iv{\frac{\tilde{j}_1^2}{R^2}} {\bf r}^2, $ 
where ${\bf r}^2=r_1^2+r_2^2+r_3^2, $ one obtain the commutation relations
for the quantum analogs of Carroll kinematics $C_v^{4(\pm)}(\sigma_0)$ with positive $(\tilde{j}_1=1)$ and negative $(\tilde{j}_1=i)$ space curvature $(i,k=1,2,3 )$
\begin{equation}
C_v^{4(\pm)}(\sigma_0)=\{t,{\bf r}|\;\;  
[t,r_k]=ivr_k(1+{\frac{\tilde{j}_1^2}{R^2}}{\bf r}^2), \;\;
 [r_i,r_k]=0\;  \}.
\label{15q}
\end{equation}
The quantum Carroll kinematics with zero curvature is achieved in the limit
 $R \rightarrow \infty $  and is as follows
\begin{equation}
C_v^{4(0)}(\sigma_0)=\{t,{\bf r}|\;\;  [t,r_k]=ivr_k, \;\;
 [r_i,r_k]=0,\; i,k=1,2,3 \}.
\label{16q}
\end{equation}
This kinematics is the non-commutative analog of the Carroll kinematics
first introduced in \cite{LL-65}.

%%%%%%%%%%%%%%%%%%%%%%%%%%%%%%%%%%%%%%%%%%%%%%%%%%%%%%
\subsection{ Quantum vector space $O^5_v(\j;\sigma')$}
%%%%%%%%%%%%%%%%%%%%%%%%%%%%%%%%%%%%%%%%%%%%%%%%%%%%%
For the permutation $\sigma'=(1,4,3,5,2)$ 
the deformation parameter is multiplied by $J=(1,2)\bigcup (4,5)=j_1j_4$
and commutation relations of the generators are  $(m=3,4,5) $
$$
\xi_{1}\xi_{m}=\xi_{m}\xi_{1} \cosh Jv
-i\xi_{m}\xi_{2} j_1 \sinh Jv , \quad
\xi_{m}\xi_{2}=\xi_{2}\xi_{m} \cosh Jv
-i\xi_{1}\xi_{m} {1 \over j_1 }\sinh Jv ,\quad
$$
$$
\xi_{4}\xi_{3}=\xi_{3}\xi_{4} \cosh Jv
-i\xi_{3}\xi_{4} j_4 \sinh Jv , \quad
\xi_{3}\xi_{5}=\xi_{5}\xi_{3} \cosh Jv
-i\xi_{2}\xi_{3} {1 \over j_4 }\sinh Jv ,
$$
$$
 [\xi_{1},\xi_{_2}] = 
2ij_1 \Biggl( j_2^2\xi_{3}^2  \cosh Jv +
( \xi_{2}^2  + j_2^2j_3^2\xi_{4}^2  )\cosh {Jv \over 2} \Biggr)
\sinh {Jv \over 2}, 
$$
\begin{equation}
[\xi_{4},\xi_{5}]=2i\xi_3^2{1 \over j_3^2j_4 }\sinh {Jv \over 2}. 
\label{17q}
\end{equation}
The  quadratic form 
$$
inv(j)  =  \Biggl( 
\xi_{1}^2  + j_1^2 \xi_{2}^2  +   
+ j_1^2j_2^2j_3^2 ( \xi_{4}^2  + j_4^2 \xi_{5}^2  ) \cosh Jv+ 
$$
\begin{equation}
 + j_1^2j_2^2\xi_{3}^2 {{(\cosh Jv)^2}\over {\cosh({Jv/2})}} 
\Biggr)\cosh {3Jv \over 2}
\label{18q}
\end{equation}
is invariant under the coaction of $SO_v(5;j;\sigma').$

For the left and right time and space generators and the physical
contraction parameters the commutation relations of the (anti) de Sitter
kinematics $S_v^{4(\pm)}(\sigma')$ are written  in the form
$$
S_v^{4(\pm)}(\sigma')=\{t,{\bf r}|\;\;
\hat{r}_kt=\hat{t}r_k \cosh {\frac{\tilde{j}_1v}{T}}
-i r_k {\frac{T}{\tilde{j}_1}} \sinh {\frac{\tilde{j}_1v}{T}},
$$
$$
\hat{r}_2r_1=\hat{r}_1r_2 \cosh {\frac{\tilde{j}_1v}{T}}
-i\hat{r}_1 r_3  \sinh {\frac{\tilde{j}_1v}{T}}, \;\;
\hat{r}_1r_3=\hat{r}_3r_1 \cosh {\frac{\tilde{j}_1v}{T}}
-i\hat{r}_2 r_1  \sinh {\frac{\tilde{j}_1v}{T}}, \;\;
$$
\begin{equation}
\hat{r}_2r_3-\hat{r}_3r_2=2i\hat{r}_1r_1\sinh {\frac{\tilde{j}_1v}{2T}}  \}.
\label{19q}
\end{equation}
The right and left generators are connected as follows
$$
\hat{r}_k=r_k \cosh {\frac{\tilde{j}_1v}{T}}
+i \hat{t}r_k {\frac{\tilde{j}_1}{T}} \sinh {\frac{\tilde{j}_1v}{T}}, \;\;
$$
\begin{equation}
\hat{t}=t+ 2i{\frac{\tilde{j}_1}{c^2T}} \left (
\hat{r}_1r_1 \cosh {\frac{\tilde{j}_1v}{T}}+
(\hat{r}_2r_2+ \hat{r}_3r_3)\cosh {\frac{\tilde{j}_1v}{2T}} \right )
\sinh {\frac{\tilde{j}_1v}{2T}}.
\label{20q}
\end{equation}
The deformation parameter $v$ has time dimension
$[v]=[T]=[\mbox{time}]=[\mbox{energy}]^{-1},$ 
therefore kinematics  $S_v^{4(\pm)}(\sigma')$ are not isomorphic
to kinematics   $S_v^{4(\pm)}(\sigma_0),\;$ (\ref{9q}).
The quantum (anti) de Sitter kinematics $S_v^{4(\pm)}(\sigma')$ may be
regarded as a kinematics with the fundamental time. 

The quantum $(1+3)$ Minkowski kinematics $M_v^4(\sigma')$ is obtained
from  the quantum (anti) de Sitter kinematics $S_v^{4(\pm)}(\sigma')$
in the zero curvature limit  $T \rightarrow \infty. $ 
The left and right generators are coincided $\hat{t}=t, \hat{r}_k=r_k$ 
in this limit and we have
\begin{equation}
M_v^4(\sigma')=\{t,{\bf r}|\;\;  [t,r_k]=ivr_k, \;  \;
[r_i,r_k]=0, \;i,k=1,2,3    \}.
\label{21q}
\end{equation}
This kinematics is isomorphic to the standard $\kappa$-Minkowski
kinematics (\ref{i2}).

In the non-relativistic limit $ c \rightarrow \infty $ the quantum kinematics $S_v^{4(\pm)}(\sigma')$ are contracted to the non-commutative
analogs of $(1+3)$ non-relativistic Newton kinematics
 $N_v^{4(\pm)}(\sigma')$ with non-zero curvature.
In this limit  the deformation parameter remain untouched, 
the right and left  time generators are the same $\hat{t}=t, $ but
the right and left space generators are related as 
\begin{equation}
\hat{r}_k=r_k \cosh {\frac{\tilde{j}_1v}{T}}
+i tr_k {\frac{\tilde{j}_1}{T}} \sinh {\frac{\tilde{j}_1v}{T}}. 
\label{22q}
\end{equation}
The quantum Newton kinematics are given by the commutation relations
$$
N_v^{4(\pm)}(\sigma')=\{t,{\bf r} | \; 
[t,r_k]=i(r_k+{\frac{\tilde{j}_1^2}{T^2}}tr_kt){\frac{T}{\tilde{j}_1}} 
\tanh{\frac{\tilde{j}_1v}{T}}, 
$$
$$
r_2r_1=r_1r_2 \cosh {\frac{\tilde{j}_1v}{T}}
-ir_1 r_3  \sinh {\frac{\tilde{j}_1v}{T}}, \;\;
r_1r_3=r_3r_1 \cosh {\frac{\tilde{j}_1v}{T}}
-ir_2 r_1  \sinh {\frac{\tilde{j}_1v}{T}}, \;\;
$$
\begin{equation}
[r_2,r_3]=2ir_1^2\sinh {\frac{\tilde{j}_1v}{2T}}  \}.
\label{23q}
\end{equation}

 In the zero curvature limit $ T \rightarrow \infty $
both quantum Newton kinematics are contracted to the quantum Galilei
kinematics
\begin{equation}
G_v^4(\sigma')=\{t,{\bf r}|\;\;  [t,r_k]=ivr_k, \;\;
 [r_i,r_k]=0, \;i,k=1,2,3 \},
\label{24q}
\end{equation}
which commutation relations are identical with those of the Minkowski kinematics (\ref{21q}). 

To obtain Carroll kinematics let us replace the parameter $j_1$
by $\tilde{j}_1R^{-1},$ where $R \rightarrow \infty $ correspond to
$j_1=\iota_1 $ and    $[R]=[\mbox{length}]. $ The parameter $j_4$
is replaced by $c,$ where $[c]=[\mbox{velocity}] $ and 
$ c \rightarrow 0 $ correspond to $j_4=\iota_4. $
Then the deformation parameter  receive the physical time dimension
$[v]=[R][c]^{-1}=[\mbox{time}]=[\mbox{energy}]^{-1},$ that is the same
dimension as for the standard kinematics.
Commutation relations (\ref{17q}) are rewritten in the form $(m=3,4,5)$
$$
\xi_{1}\xi_{m}=\xi_{m}\xi_{1} \cosh {\frac{\tilde{j}_1cv}{R}}
-i \xi_m\xi_2 {\frac{\tilde{j}_1}{R}} \sinh {\frac{\tilde{j}_1cv}{R}}, \;\;
$$
$$
\xi_{m}\xi_{2}=\xi_{2}\xi_{m} \cosh {\frac{\tilde{j}_1cv}{R}}
-i \xi_1\xi_m {\frac{R}{\tilde{j}_1}} \sinh {\frac{\tilde{j}_1cv}{R}},\;\;
$$
$$
\xi_{4}\xi_{3}=\xi_{3}\xi_{4} \cosh {\frac{\tilde{j}_1cv}{R}}
-ic \xi_3\xi_5  \sinh {\frac{\tilde{j}_1cv}{R}},\;\;
$$
$$
\xi_{3}\xi_{5}=\xi_{5}\xi_{3} \cosh {\frac{\tilde{j}_1cv}{R}}
-i \xi_4\xi_3 {\frac{1}{c}} \sinh {\frac{\tilde{j}_1cv}{R}},\;\;
$$
$$
[\xi_{1},\xi_{2}]=2i {\frac{\tilde{j}_1}{R}} \left (
\xi_3^2 \cosh {\frac{\tilde{j}_1cv}{R}}+
(\xi_4^2+c^2\xi_5^2)\cosh {\frac{\tilde{j}_1cv}{2R}} \right )
\sinh {\frac{\tilde{j}_1cv}{2R}},
$$
\begin{equation}
[\xi_{4},\xi_{5}]= 2i\xi_3^2
 {\frac{1}{c}} \sinh {\frac{\tilde{j}_1cv}{2R}}
\label{25q}
\end{equation}
and in the limit  $ c \rightarrow 0 $ are as follows
$$
[\xi_{1},\xi_{m}]=0,\;\; [\xi_{2},\xi_{m}]=0,\;\;[\xi_{1},\xi_{2}]=0,\;\;
[\xi_{3},\xi_{4}]=0,
$$
\begin{equation}
\xi_{3}\xi_{5}=\xi_{5}\xi_{3}-iv {\frac{\tilde{j}_1}{R}} \xi_{4}\xi_{3},\;\;
[\xi_{4},\xi_{5}]=iv {\frac{\tilde{j}_1}{R}} \xi_{3}^2.
\label{26q}
\end{equation}

The generator $\xi_1$ commute with $\xi_s,\;s=2,3,4,5,$ therefore
the left and right generators are coincided
 $\hat{r}_k=r_k=\xi_{k+1} \xi_1^{-1}, \; k=1,2,3,\; 
\hat{t}=t=\xi_{4} \xi_1^{-1}$ 
and commutation relations of the quantum Carroll kinematics
$C_v^{4(\pm)}(\sigma')$ are easily obtained $(i,k=1,2,3)$
$$
C_v^{4(\pm)}(\sigma')=\{t,{\bf r}|\;\;[t,r_1]=0,\;\;  
[t,r_2]=iv{\frac{\tilde{j}_1}{R}}r_3r_2, \;\; 
$$
\begin{equation}
[r_3,t]=iv{\frac{\tilde{j}_1}{R}}r_1^2,\;\;
 [r_i,r_k]=0  \}.
\label{27q}
\end{equation}
As before the quantum Carroll kinematics with zero curvature is achieved in the limit $R \rightarrow \infty $  
\begin{equation}
C_v^{4(0)}(\sigma')=\{t,{\bf r}|\;\;[t,r_k]=0,\;\;  
 [r_i,r_k]=0,\;i,k=1,2,3  \}.
\label{27q'}
\end{equation}
It is remarkable that all commutators are equal to zero and the quantum
Carroll kinematics is identical with the commutative Carroll kinematics
\cite{LL-65}.
%%%%%%%%%%%%%%%%%%%%%%%%%%%%%%%%%%%%%%%%%%%%%%%%%%%%%%
\subsection{ Quantum vector space $O^5_v(\j;\tilde{\sigma})$}
 
%%%%%%%%%%%%%%%%%%%%%%%%%%%%%%%%%%%%%%%%%%%%%%%%%%%%%
For the permutation  $\tilde{\sigma}=(2,3,1,4,5)$
the deformation parameter is multiplied by $J=(2,5)\bigcup (3,4)=j_2j_3j_4$
and commutation relations of the generators are $(m=1,3,4)$ 
$$
\xi_{2}\xi_{m}=\xi_{m}\xi_{2} \cosh Jv
-i\xi_{m}\xi_{5} J \sinh Jv , \;\;
\xi_{5}\xi_{m}=\xi_{m}\xi_{5} \cosh Jv
+i\xi_{m}\xi_{2} {1 \over J }\sinh Jv ,
$$
$$
\xi_{3}\xi_{1}=\xi_{1}\xi_{3} \cosh Jv
-i\xi_{1}\xi_{4} j_3 \sinh Jv , \;\;
\xi_{4}\xi_{1}=\xi_{1}\xi_{4} \cosh Jv
+i\xi_{1}\xi_{3} {1 \over j_3 }\sinh Jv ,
$$
$$
 [\xi_{2},\xi_{_5}] = 
2i{1 \over j_1^2} \left ( \xi_{1}^2  \cosh(Jv) +
j_1^2j_2^2( \xi_{3}^2  + j_3^2\xi_{4}^2  )\cosh {Jv \over 2} \right )
{ 1 \over J}\sinh {Jv \over 2} .
$$
\begin{equation}
[\xi_{3},\xi_{4}]=2i\xi_1^2{1 \over j_1^2j_2^2j_3 }\sinh{Jv \over 2}. 
\label{28q}
\end{equation}
The  quadratic form
$$
inv(j)  = \Biggl( \xi_{1}^2 {{(\cosh(Jv))^2}\over {\cosh({Jv/2})}} 
+j_1^2(\xi_{2}^2  + J^2 \xi_{5}^2)  +   
$$
\begin{equation}
+ j_1^2j_2^2( \xi_{3}^2 + j_3^2 \xi_{4}^2)\cosh (Jv) \Biggr) \cosh {3Jv \over 2}
\label{29q}
\end{equation}
is invariant under the coaction (\ref{9}) of $SO_v(5;j; \tilde{\sigma}).$

For the left and right time and space generators and the physical
contraction parameters the commutation relations of the (anti) de Sitter
kinematics  $S_v^{4(\pm)}(\tilde{\sigma})$ are written  in the form
$(p=1,2)$
$$
S_v^{4(\pm)}(\tilde{\sigma})=\{t,{\bf r}|\;\;
\hat{t}r_p=\hat{r}_pt \cos {\frac{v}{c}}
+i\hat{r}_pr_3 {\frac{1}{c}} \sin {\frac{v}{c}},
$$
$$
\hat{t}r_3-\hat{r}_3t=2i{\frac{cT^2}{\tilde{j}_1^2}} \Biggl( 
\cos {\frac{v}{c}} - {\frac{\tilde{j}_1^2}{c^2T^2}}(\hat{r}_1r_1+\hat{r}_2r_2)
\cos {\frac{v}{2c}} \Biggr)\sin {\frac{v}{2c}},
$$
\begin{equation}
\hat{r}_pr_3=\hat{r}_3r_p \cos {\frac{v}{c}} 
-i\hat{t} r_p  c \sin {\frac{v}{c}}, \;\;
\hat{r}_1r_2-\hat{r}_2r_1=2{\frac{c^2T^2}{\tilde{j}_1^2}} \sin {\frac{v}{2c}}
  \}.
\label{30q}
\end{equation}
The left and right generators are connected by the  following relations
$$
\hat{t}=t\cos {\frac{v}{c}}  + i r_3 {\frac{1}{c}} \sin {\frac{v}{c}}, \;\;
\hat{r}_1=r_1\cos {\frac{v}{c}}  +  r_2  \sin {\frac{v}{c}},
$$
\begin{equation}
\hat{r}_2=r_2\cos {\frac{v}{c}}  -  r_1  \sin {\frac{v}{c}},\;\;
\hat{r}_3=r_3\cos {\frac{v}{c}}  + i t c  \sin {\frac{v}{c}}.
\label{31q}
\end{equation}
The deformation parameter  has the velocity dimension
$[v]=[c]=[\mbox{velocity}],$ 
therefore kinematics $S_v^{4(\pm)}(\tilde{\sigma})$  are not isomorphic
to kinematics   $S_v^{4(\pm)}(\sigma_0),\;$ (\ref{9q}) and
$S_v^{4(\pm)}(\sigma'),$ (\ref{19q}) and may be
regarded as a kinematics with the fundamental velocity. 
As it follows from (\ref{31q}), both contractions $T \rightarrow \infty, $
$ c \rightarrow \infty $ are not  permitted, therefore the quantum (anti) de Sitter kinematics $S_v^{4(\pm)}(\tilde{\sigma})$ have not Minkowski, Newton, Galilei kinematics as a limiting cases.

To obtain Carroll kinematics let us replace the parameter $j_1$
by $\tilde{j}_1R^{-1},$ where $R \rightarrow \infty $ correspond to
$j_1=\iota_1 $ and    $[R]=[\mbox{length}]. $ The parameter $j_4$
is replaced by $c,$ where $[c]=[\mbox{velocity}] $ and 
$ c \rightarrow 0 $ correspond to $j_4=\iota_4. $
Then the deformation parameter  receive the physical  dimension
$[v]=[c]^{-1}=[\mbox{velocity}]^{-1},$
that is the inverse dimension as compare with  the standard kinematics.
Commutation relations (\ref{28q}) are rewritten in the form $(m=1,3,4)$
$$
\xi_{2}\xi_{m}=\xi_{m}\xi_{2} \cosh (cv) -ic \xi_m\xi_5  \sinh (cv), \;\;
\xi_{5}\xi_{m}=\xi_{m}\xi_{5} \cosh (cv) +i \xi_m\xi_2 {\frac{1}{c}} \sinh (cv),
$$
$$
\xi_{3}\xi_{1}=\xi_{1}\xi_{3} \cosh (cv) -i \xi_1\xi_4  \sinh (cv), \;\;
\xi_{4}\xi_{1}=\xi_{1}\xi_{4} \cosh (cv) +i \xi_1\xi_3  \sinh (cv),
$$
$$
[\xi_{2},\xi_{5}]=2i {\frac{R^2}{c\tilde{j}_1^2}} \left (
\xi_1^2 \cosh (cv) + {\frac{\tilde{j_1^2}}{R^2}}
(\xi_3^2 + \xi_4^2)\cosh {\frac{cv}{2}} \right ) \sinh {\frac{cv}{2}},
$$
\begin{equation}
[\xi_{3},\xi_{4}]= 2i {\frac{R^2}{\tilde{j}_1^2}} \sinh {\frac{cv}{2}}
\label{32q}
\end{equation}
and in the limit  $ c \rightarrow 0 $ are
$$
[\xi_{2},\xi_{m}]=0,\;\; [\xi_{1},\xi_{3}]=0,\;\;[\xi_{1},\xi_{4}]=0,\;\;
[\xi_{3},\xi_{4}]=0,
$$
\begin{equation}
\xi_{5}\xi_{m}=\xi_{m}\xi_{5}+iv  \xi_{m}\xi_{2},\;\;
[\xi_{2},\xi_{5}]=iv {\frac{R^2}{\tilde{j_1^2}}} \left (
\xi_{1}^2 + {\frac{\tilde{j_1^2}}{R^2}}(\xi_3^2 + \xi_4^2) \right ).
\label{33q}
\end{equation}
As far as $\xi_1$ commute with $\xi_2,\xi_3,\xi_4,$ the left and right
space generators are equal
$\hat{r}_k=r_k=\xi_{k+1} \xi_1^{-1}, \; k=1,2,3, $
but the left and right time generators are related as
$\hat{t}=t+ivr_1$ and commutation relations of the quantum Carroll kinematics $C_v^{4(\pm)}(\tilde{\sigma})$ with curvature are as follows
 $(i,k=1,2,3,\;p=2,3)$
\begin{equation}
C_v^{4(\pm)}(\tilde{\sigma})=\{t,{\bf r}|\;\;[t,r_p]=0,\;\;  [r_i,r_k]=0, \;\;
[r_1,t]=iv \left ( {\frac{R^2}{\tilde{j_1^2}}} + {\bf r}^2 \right ) \}.
\label{34q}
\end{equation}
There is no quantum analogs of the flat Carroll kinematics \cite{LL-65},
since the limit $R \rightarrow \infty $ is forbidden.

%%%%%%%%%%%%%%%%%%%%%%%%%%%%%%%%%%%%%%%%%%%%%%%%%%%%%%
\section{Conclusion}
%%%%%%%%%%%%%%%%%%%%%%%%%%%%%%%%%%%%%%%%%%%%%%%%%%%%%%

The narrow analysis of the multiplier
$J=(\sigma_1,\sigma_5)\cup (\sigma_2,\sigma_4),$
which is appeared in the transformation of the deformation parameter
$z=Jv,$ and commutation relations (\ref{1q}) of the quantum vector space generators for different permutations $ \sigma $ made possible to find
three permutations giving a different $J$ and a physically nonequivalent
kinematics. For the identical permutation $ \sigma_0 $ the quantum (anti) de Sitter kinematics (\ref{9q}) are characterized
by the fundamental length $[v]=[\mbox{length}], $ 
for the  permutation $ \sigma' $   are characterized
by the fundamental time $[v]=[\mbox{time}] $ (\ref{19q}) and
for the  permutation $ \tilde{\sigma} $ ---   
by the fundamental velocity $[v]=[\mbox{velocity}] $ (\ref{30q}).
Recall that the same physical dimensions of the deformation parameter
have been received for the quantum algebras $so_v(3;j;\sigma)$ and corresponding $(1+1)$ kinematics for a different permutations \cite{G-95}.

In the zero curvature limit $ T \rightarrow \infty $ two quantum
Minkowski kinematics (\ref{10q}) and (\ref{21q}) have been obtained 
$$
M_v^4(\sigma_0)=\{t,{\bf r}|\;\;  [t,r_p]=0, \;[r_3,t]=ivt, \; 
$$
$$
[r_2,r_1]=0,\; [r_3,r_p]=ivr_p, \;p=1,2,    \},
$$
\begin{equation}
M_v^4(\sigma')=\{t,{\bf r}|\;\;  [t,r_k]=ivr_k, \;  \;
[r_i,r_k]=0, \;i,k=1,2,3    \}.
\label{35q}
\end{equation}
The first one is isomorphic to the tachyonic $\kappa$-deformation (\ref{i3}),
the second one to the standard  $\kappa$-deformation (\ref{i2}).
For both $\kappa$-Minkowski kinematics (\ref{i2}),(\ref{i3}) in the system units $\hbar=c=1$ the deformation parameter $\Lambda=\kappa^{-1}$ has
the physical dimension of length and is interpreted as the fundamental
length. But in the system units $\hbar=1$ the deformation parameter
has a different dimensions, namely $v$ is the fundamental length for
$M_v^4(\sigma_0)$ kinematics and $v$ is the fundamental time for
$M_v^4(\sigma'). $

As far as the commutation relations (\ref{35q}) do not depend on $c$
they are not changed in the limit $ c \rightarrow \infty,$ therefore
the generators of the quantum Galilei kinematics $G_v^4(\sigma_0)$ (\ref{12q}) and $G_v^4(\sigma')$ (\ref{23q}) are subject of the same
commutation relations. The only difference consist in following:
for the  Galilei kinematics there are two invariants $inv_1=t^2,  inv_2=r_1^2+r_2^2 +r_3^2 $ with respect to the coaction of the corresponding quantum groups, whereas for the Minkowski kinematics
there is only one invariant $inv=t^2-(r_1^2+r_2^2+r_3^2 ).$
Thereby the quantum deformations of the flat kinematics are identical
up to the coaction of the corresponding quantum groups
for both  relativistic and non-relativistic one. It is remarkable that
the same invariants are in the commutative case.

There are two non-commutative analogs of the non-relativistic Newton
kinematics (\ref{11q}),  (\ref{23q})
$$
N_v^{4(\pm)}(\sigma_0)=\{t,{\bf r} | \; [t,r_p]=0, \;\;
[r_3,t]=ivt(1+\tilde{j}_1^2{\frac{t^2}{T^2}}), \;\;
$$        
$$
[r_1,r_2]=0,\;\;
[r_3,r_p]=ivr_p(1+\tilde{j}_1^2{\frac{t^2}{T^2}}),\; p=1,2 \}, 
$$
$$
N_v^{4(\pm)}(\sigma')=\{t,{\bf r} | \; 
[t,r_k]=i(r_k+{\frac{\tilde{j}_1^2}{T^2}}tr_kt){\frac{T}{\tilde{j}_1}} 
\tanh{\frac{\tilde{j}_1v}{T}}, 
$$
$$
r_2r_1=r_1r_2 \cosh {\frac{\tilde{j}_1v}{T}}
-ir_1 r_3  \sinh {\frac{\tilde{j}_1v}{T}}, \;\;
r_1r_3=r_3r_1 \cosh {\frac{\tilde{j}_1v}{T}}
-ir_2 r_1  \sinh {\frac{\tilde{j}_1v}{T}}, \;\;
$$
\begin{equation}
[r_2,r_3]=2ir_1^2\sinh {\frac{\tilde{j}_1v}{2T}}  \},
\label{36q}
\end{equation}
where in the last case the deformation parameter do not transformed
under contraction. The multiplier $T^{-1}$ is appeared as the result of
the physical interpretations of the quantum space generators.

There are three non-commutative analogs of the exotic non-zero curvature Carroll kinematics (\ref{15q}), (\ref{27q}),  (\ref{34q})
$$
C_v^{4(\pm)}(\sigma_0)=\{t,{\bf r}|\;\;  
[t,r_k]=ivr_k(1+{\frac{\tilde{j}_1^2}{R^2}}{\bf r}^2), \;\;
 [r_i,r_k]=0,\;  \},
$$
$$
C_v^{4(\pm)}(\sigma')=\{t,{\bf r}|\;\;[t,r_1]=0,\;\;  
[t,r_2]=iv{\frac{\tilde{j}_1}{R}}r_3r_2, \;\; 
$$
$$
[r_3,t]=iv{\frac{\tilde{j}_1}{R}}r_1^2,\;\;
 [r_i,r_k]=0,  \},
$$
\begin{equation}
C_v^{4(\pm)}(\tilde{\sigma})=\{t,{\bf r}|\;\;[t,r_p]=0,\;\;  [r_i,r_k]=0, \;\;
[r_1,t]=iv \left ( {\frac{R^2}{\tilde{j_1^2}}} + {\bf r}^2 \right ) \}
\label{37q}
\end{equation}
and two quantum analogs of the  zero curvature Carroll kinematics 
(\ref{16q}), (\ref{27q'})
$$
C_v^{4(0)}(\sigma_0)=\{t,{\bf r}|\;\;  [t,r_k]=ivr_k, \;\;
 [r_i,r_k]=0,\; i,k=1,2,3 \},
$$
\begin{equation}
C_v^{4(0)}(\sigma')=\{t,{\bf r}|\;\;[t,r_k]=0,\;\;  
 [r_i,r_k]=0,\;i,k=1,2,3  \}.
\label{38q'}
\end{equation}
The deformation parameter has the physical dimension of time
 $[v]=[\mbox{time}]$ for permutations  $\sigma_0, \sigma' $
and the  dimension of inverse velocity
 $[v]=[\mbox{velocity}]^{-1}$ for permutation  $\tilde{\sigma}.  $

In spite of the fact that  the commutation relations of generators of
$C_v^{4(0)}(\sigma_0) $ and $M_v^4(\sigma') $ are identical, both kinematics are physically different. Mathematically isomorphic kinematics may be physically non-equivalent.
 
\section{Acknowledgments}

One of the authors (NG) thank J.Lukierski for pointing out the papers on
$\kappa$-Minkowski.


\begin{thebibliography}{99}
%
\bibitem{B-68}
Bacry H and Levy--Leblond J-M 1968 {\it J.Math.Phys}
 {\bf 9}  1605
%
\bibitem{S-47}
 Snyder H S 1947 {\it Phys.Rev.} {\bf 71}  38
%
\bibitem{K-61} 
Kadyshevsky V G 1961 {\it  Doclady AS USSR} {\bf 136} 70 (in Russian) 
%
\bibitem{K-62} 
Kadyshevsky V G 1962 {\it  Doclady AS USSR} {\bf 147} 588 (in Russian) 
%
\bibitem{G-62}
Golfand Yu A 1962 {\it Sov. Phys. JETP} {\bf 43} 256 (in Russian) 
%
\bibitem{T-65}
Tamm I E  1975 {\it Collection of Scientific Works} (Moscow: Nauka)
{\bf 2} 218 (in Russian)  
%
\bibitem{TV-72}
Tamm I E and Vologodski V B 1975 {\it Collection of Scientific Works} (Moscow: Nauka) {\bf 2} 226 (in Russian) 
%
\bibitem{MK-96}
Mir-Kasimov R M 1996 {\it Phys. Lett. B }  {\bf 378} 181 
%
\bibitem{FRT} 
Faddeev L D, Reshetikhin N Yu and Takhtajan L A  1989 {\it Algebra i Analis} {\bf 1} 178 (in Russian) 
%
\bibitem{Z-94}
Zakrzewski S  {\it J. Phys. A: Math. Gen.} {\bf 27} 2075
%
\bibitem{KM-95}
Kosinski P and  Maslanka P 1996 {\it From field theory to quantum groups} (World Scientific) 41; {\it Preprint} q-alg/9512018
%
\bibitem{LLM}
Lukierski J,  Lyakhovsky V and  Mozrzymas M {\it Preprint}
hep-th/0203182
%
\bibitem{BHOS-96}
Ballesteros A,  Herranz F J,  del Olmo M A and  Santander M 1996
{\it Phys. Lett.} {\bf B351} 215
%
\bibitem{GYa-86}
Gromov N A and Yakushevich L V 1986 {\it Group-theoretical methods in Physics} (Moscow: Nauka) {\bf 2} 191 (in Russian)
%
\bibitem{G-90}
Gromov N A 1990 {\it Contractions and Analytical Continuations of
Classical Groups. Unified Approach} (Syktyvkar: Komi SC) (in Russian)
%
\bibitem{GKK-97} 
Gromov N A , Kostyakov I V and Kuratov V V  1997
Quantum orthogonal Cayley-Klein groups in Cartesian  basis.
{\it Int.J.Mod.Phys.A} {\bf 12} 33; {\it Preprint} q-alg/9610011
%
\bibitem{G-97} 
Gromov N A , Kostyakov I V and Kuratov V V  1997
{\it Quantum Group Symposium at Group21} (Sofia: Heron Press ) 202
%
\bibitem{Sb-97} 
Gromov N A , Kostyakov I V and Kuratov V V  1997
{\it Algebra, Differential Equations and Probability Theory}
(Syktyvkar: Komi SC) 3 (in Russian)
%
\bibitem{GKK-98} 
Gromov N A , Kostyakov I V and Kuratov V V  1998
{\it Proc.  5th Wigner Symp. (Vienna, Austria, 
25-29 August 1997)} (Singapore: World Scientific)
 19; {\it Preprint} q-alg/9710009
%
\bibitem{GK-97}
Gromov N A , Kostyakov I V and Kuratov V V  1998
 {\it Preprint} q-alg/9711024
%
\bibitem{BHOS-93}
Ballesteros A,  Herranz F J,  del Olmo M A and  Santander M 1993
{\it J. Phys. A: Math. Gen. } {\bf 26} 5801
%
\bibitem{BHOS-95}
Ballesteros A,  Herranz F J,  del Olmo M A and  Santander M 1995
{\it  Lett. Math. Phys.} {\bf 33} 273
%
\bibitem{G-95} 
Gromov N A 1997  
{\it Turkish J. Phys.} {\bf 3} 377 {\it Preprint} q-alg/9602003
%
\bibitem{Sb-2000} 
Gromov N A , Kostyakov I V and Kuratov V V  2000
{\it Algebra, Differential Equations and Probability Theory}
(Syktyvkar: Komi SC) 3 (in Russian)
%
\bibitem{GKK-2001} 
Gromov N A , Kostyakov I V and Kuratov V V  2001
{\it Phys. Atom. Nucl.} {\bf 64} 1963; {\it Preprint} math.QA/0102071
%
\bibitem{GKK-2002} 
Gromov N A , Kostyakov I V and Kuratov V V  2002
{\it Preprint} math.QA/0209158
%
\bibitem{Sb-2003}
Gromov N A , Kostyakov I V and Kuratov V V  2003
{\it Algebra, Geometry and Differential Equations }
(Syktyvkar: Komi SC) 4 (in Russian)
%
\bibitem{LL-65}
Levy-Leblond J-M 1965
{\it Ann. Inst. H.Poincar\'e} {\bf A3} 1
%
\end{thebibliography}
\end{document}